# SEARCH FOR A SIGNAL ON QCD CRITICAL POINT IN CENTRAL NUCLEUS-NUCLEUS COLLISIONS

M. K. Suleymanov[1,2], E. U. Khan[1], K. Ahmed[1], Mahnaz Q. Haseeb[1], Farida Tahir[1], Y. H. Huseynaliyev[1], M. Ajaz[1], K. H. Khan[1], Z.Wazir[1]

[1]Dept. of Physics
CIIT
Islamabad, Pakistan
[2]VBLHE
JINR
Dubna, Russia.

**Abstract** – We discuss that the QCD critical point could appear in central collisions in percolation cluster. We suggest using the nuclear transparency effect and the one of the light nuclear production to identify the critical point.

## INTRODUCTION

During the last several years we are discussing some results of the central experiments (see for example [1]). These demonstrate the point of regime change and saturation on the behavior of some characteristics of the events as a function of the centrality. We believe that such phenomena could be connected with fundamental properties of the strongly interacting mater and could reflect the changes of its states (phases). The most efficient and cost effective way to get the information on these properties and the phases of strongly interacting matter is to use the central experiments.

BNL E910 has measured $\Lambda$ production as a function of collision centrality for 17.5 GeV/c p–Au reactions [2]. It was observed that the measured $\Lambda$ yield increases with centrality faster then saturates. This Collaboration has obtained the same results for $K^0_s$ and $K^+$-mesons emitted in p+Au reaction.

In Ref. [3] the variations of average transverse mass of identified hadrons with charge multiplicity have been studied for AGS, SPS and RHIC energies (Fig.1). A plateau was observed in the average transverse mass for multiplicities corresponding to SPS energies. It was claimed that it can be attributed to the formation of a co-existence phase of quark gluon plasma and hadrons.

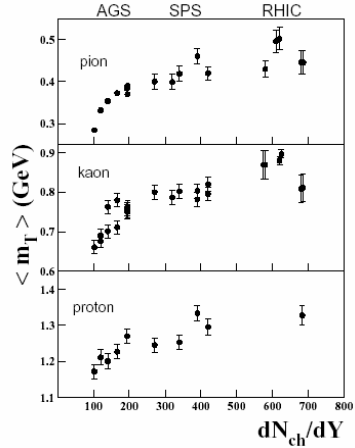

Figure 1 : Variation of $<m_T>$ with produced charged particles per unit rapidity at mid rapidity for central collisions corresponding to different $\sqrt{s}$ spanning from AGS to RHIC.

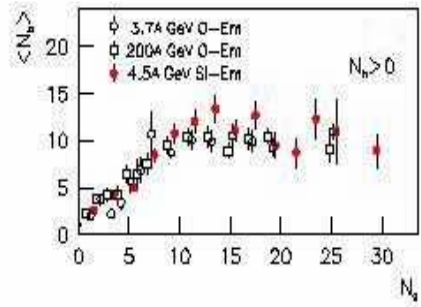

Figure 2 : $N_g$ –dependences of $<N_b>$ for different reactions.

Fig. 2 is a plot of multiplicity of grey particles - $N_g$–dependences versus $<N_b>$ average multiplicity of b-particles for different reactions taken from [4]. One can see that the values of $<N_b>$ increase with $N_g$ in the region of the values of $N_g < 8$. Then the values of the $<N_b>$ saturate in the region $N_g \geq 8$ as well as in [5].

## MAIN RESULTS

We consider these phenomena in a way that: the point indicates to critical phenomenon in wide incident energy and colliding mass range independent of the type of secondary particles; in some cases after the point of regime change saturation is observed.

These results could be interpreted as: the phenomenon has scale invariant behavior which may mean that the parton degrees of freedom are responsible for it;

which does not depend on quark flavours. These behaviors could be expected around the critical point.

Before continuing our discussion let us comment that the definition of the centrality is not simple problem because it cannot be defined directly in the experiment. In different experiments the values of the centrality are fixed by different ways. Apparently, it is not simple to compare quantitatively the results on centrality-dependences obtained in literature while on the other hand the definition of centrality could significantly influence the final results.

## EXPLANATION OF THE RESULTS

If the regime change takes place unambiguously two times, this would be surely the most direct experimental evidence seen to observe the QCD critical point and phase transition. But the central experiments could not confirm it. It may be due to the appearance of critical transparency of the matter nears the critical point. Because in high density matter, near the critical point, quarks and partons could be bound as a result of the percolation and the matter might behave as a superconductor [6]. We therefore think on percolation cluster that to explain the result it is necessary to suggest that the dynamics is the same for hadron-nucleus and nucleus –nucleus collisions, independent of energy and mass of the colliding nuclei and their types. The mechanism to describe the phenomena may be statistical or percolative due to their critical character [7]. We believe that the mechanism to explain the phenomena may be the percolation cluster formation [8], which makes a multibaryon system. Big percolation clusters may be formed in these interactions independent of the colliding energy but the structure, maximum density and temperature of hadronic matter may depend on colliding energy and masses in the cluster framework. The deconfinement is expected when the density of quarks and gluons becomes so high that due to strong overlap, it no longer makes sense to partition them into color-neutral hadrons [6]. The clusters get much larger than hadrons, within which color is not confined; deconfinement is thus related to cluster formation and a connection between it and percolation seems very likely [6], [9].

## SUGGESTIONS FOR INVESTIGATION

We believe that the study of the behavior of transparency function R at different energies as a function of the centrality could give the information on the onset state of the deconfinement in cluster. Using some statistical and percolation models [7] and experimental data on the behavior of R it is possible to get information on the appearance of the anomalous nuclear transparency.

Nuclear transparency is one of the effects of nuclear-nuclear collision from which one may get the information about the structure, states, properties and phases of the nuclear matter. A promising observable to map this transition is the transparency of the nuclear medium to the propagation of hadrons. Transparency depends upon different factors of the collisions.

Investigation of the light nuclei production as a function of the centrality could give the clue on freeze-out state of QGP formation, which could be used as additional information to confirm the percolation cluster formation near the critical point. There are two types of light nuclei emitted in heavy ion collisions: first type are light nuclei which get produced as a result of nucleus disintegration of the colliding nuclei; while the second one are light nuclei which are get made of protons and neutrons (for example as a result of coalescence mechanism) which were produced in heavy ion interaction. In an experiment we can separate these two types of nuclei from each other using the following ideas: the yields for first type nuclei have to decrease by some regularity with centrality of collisions. On the other hand, formation of the clusters could be a reason of the regime change of the behavior of light nuclei yields as a function of the centrality in the second type.

## CONCLUSION

Study of the nuclear transparency effect and one of light nuclei production as a function of centrality could give an important information on the QCD critical point in the framework of the percolation cluster.